\documentstyle[12pt] {article}

\begin {document}
\parindent=15pt
\begin{center}
\vskip 1.5 truecm
{\huge \bf HARD PROCESSES ON}\\
\vspace{.3cm}
{\huge \bf NUCLEAR TARGETS IN QCD}\\
\vspace{1.0cm}
N.Armesto, C.Pajares and C.A.Salgado \\
\vspace{.2cm}
{\it Departamento de F\'{\i}sica de Part\'{\i}culas, Universidade de Santiago de
Compostela, \\
15706-Santiago de Compostela, Spain} \\
\vspace{.7cm}
and\\
\vspace{.7cm}
Yu.M.Shabelski \\
\vspace{.2cm}
{\it Petersburg Nuclear Physics Institute, \\
Gatchina, St. Petersburg 188350, Russia} \\
\end{center}
\vspace{1cm}
\begin{abstract}
We calculate the A-dependence of open charm and beauty as well as
Drell-Yan pair production cross sections on nuclear targets taking into
account the difference of quark and gluon distributions in free and bound
nucleons. The results are presented for both proton-nucleus and
nucleus-nucleus collisions. For the case of heavy flavour production at
comparatively low energies, if $\sigma \sim A^{\alpha}$, the values of
$\alpha$ are slightly higher than unity.
With increasing energy $\alpha$ decreases in all cases and becomes smaller than
unity.
We also show $\alpha$ as a function of
Feynman-$x$ of the produced $Q\overline{Q}$ or $l^+l^-$ pair and obtain that
it decreases significantly in the beam fragmentation region. The sensitivity
of our results to the used set of parton distributions is also discussed.
\end{abstract}
\vspace{4.0cm}



\noindent{\large \bf September 1996}

\noindent{\large \bf hep-ph/9609296}

\noindent{\large \bf US-FT/37-96}

\newpage

\section{INTRODUCTION}

\vskip 2mm

Processes of heavy quark and lepton pair
production on nucleon and nuclear targets at
high energies are very interesting from both theoretical and practical
reasons. These processes provide a method to study the internal
structure of hadrons in the very small $x$-region. Some limits on the possible
non-perturbative contributions to this region
can also be obtained. Realistic estimates of
the cross sections of heavy quark production are necessary in order to
plan experiments on the existing and future accelerators.

Predictions are usually obtained in the framework of perturbative QCD in the 
leading and the next-to-leading order $\alpha_s$-expansion. In the case of a 
nuclear target it is usually assumed that the cross section of heavy flavour 
production (or any other hard process), $\sigma(Q\overline{Q})$, should be 
proportional to $A^{\alpha}$, with $\alpha = 1$; that is, in agreement with the
most accurate experimental results of [1,2], where values of 
$\alpha = 1.00 \pm 0.05$ for all $D$-meson production by pions at 
$\sqrt{s} = 22$ GeV and $\alpha = 1.02 \pm 0.03 \pm 0.02$ for neutral 
$D$-meson production by protons at $\sqrt{s} = 39$ GeV were obtained,
respectively.

However it seems important to consider the problem in more detail
because, for example, many experimental results are obtained by
extrapolation of results on a nuclear target to a nucleon target. It is
experimentally well-known [3-9] that parton distributions in nuclei (i.e., in
bound nucleons) are slightly different from the same distributions in free
nucleons. So the value of $\alpha$ can differ from unity and it is
interesting to estimate the size of this effect. Also the differences in the
distributions of gluons, sea quarks and valence quarks are not the same
\cite{Kari}. Moreover, in different hard processes different QCD
subprocesses give the main contributions, so the effects on the
$\alpha$-values can also be different.

In this paper we calculate the values of $\alpha$ as a function of
energy and Feynman-$x$ for the cases of open charm, beauty and
different mass Drell-Yan pair production in proton-nucleus and
nucleus-nucleus collisions.

In our previous paper \cite{APSS} we showed that the results for the
A-dependence of heavy flavour production depend very weakly
on the used set of parton distributions. Three different sets,
namely MRS-1 \cite{mrs},  MT S-DIS \cite{mt} and GRV HO \cite{grv}, which can
be found in CERN PDFLIB \cite{pdf}, gave practically the same results for the
$\alpha$ behaviour (see Table 1 of Ref. \cite{APSS}). Here we present
the results for two sets, GRV HO \cite{grv} and MRS-A \cite{mrsa} which are
in comparatively good agreement with the new HERA data and we see that
they give practically identical results for both open heavy flavour
and Drell-Yan pair production in proton-nucleus and nucleus-nucleus
collisions.

\vskip 9 mm

\section{CROSS SECTIONS OF HARD PROCESSES IN QCD}

\vskip 2mm

The standard QCD expression for the cross section of heavy quark production
in a hadron 1 - hadron 2 collision has the form

\begin{equation}
\sigma^{12\rightarrow Q\overline{Q}} = \int_{x_{a0}}^{1} \frac{dx_a}{x_a}
\int_{x_{b0}}^{1} \frac{dx_b}{x_b}\left [x_aG_{a/1}(x_a,\mu^{2})
\right ]\left [x_bG_{b/2}(x_b,\mu^{2})\right ]
\hat{\sigma}^{ab\rightarrow Q\overline{Q}}(\hat{s},m_Q,\mu^{2}) \;,
\label{eq:totalpf}
\end{equation}
where $x_{a0} = \textstyle 4m_Q^2/ \textstyle s$ and  $x_{b0} = \textstyle
4m_Q^2/ \textstyle (sx_a)$. Here $G_{a/1}(x_a,\mu^{2})$ and
$G_{b/2}(x_b,\mu^{2})$ are the structure functions of partons $a$ and $b$
inside hadrons $1$ and $2$ respectively, and $\hat{\sigma}^{ab\rightarrow
Q\overline{Q}}(\hat{s},m_Q,\mu^{2})$ is the cross section of the subprocess
$ab\rightarrow Q\overline{Q}$ as given by standard QCD. The latter depends
on the parton center-of-mass energy \mbox{$\hat{s} = (p_a+p_b)^2 =
x_ax_bs$}, the mass of the produced heavy quark $m_Q$
and the QCD scale $\mu^{2}$. Eq.  (\ref{eq:totalpf}) should
account for all possible subprocesses $ab \rightarrow Q\overline{Q}$.  The
parton-parton cross section $\hat{\sigma}^{ab\rightarrow
Q\overline{Q}}$ can be written in the form \cite{NDE}

\begin{equation}
\hat{\sigma}^{ab\rightarrow Q\overline{Q}}(\hat{s},m_Q,\mu^{2}) =
\frac{\alpha^{2}_{s}(\mu^{2})}{m^{2}_{Q}}f_{ab}(\rho ,\mu^{2},m^{2}_{Q}) \;,
\end{equation}
with

\begin{equation}
\rho = 4m^{2}_{Q}/\hat{s}
\end{equation}
and

\begin{equation}
f_{ab}(\rho, \mu^2, m^2_Q) = f^{(0)}_{ab}(\rho) + 4\pi
\alpha_{s}(\mu^{2})[f^{(1)}_{ab}(\rho) +
\hat{f}^{(1)}_{ab}(\rho)\ln{(\mu^{2}/m^{2}_{Q})}]\;.
\end{equation}

The functions $f^{(0)}_{ab}(\rho)$, $f^{(1)}_{ab}(\rho)$ and
$\hat{f}^{(1)}_{ab}(\rho)$ can be found in \cite{NDE} for heavy flavour
production.

Formulae with the same structure as
(1)-(4) can be found in \cite{DY} for Drell-Yan pair
production. In this case $M_{l^+l^-}$ plays the role of $2m_Q$ in
$x_{a0}$, $x_{b0}$ and $\hat{\sigma}^{ab\rightarrow Q\overline{Q}}$.

For the numerical calculations we wrote the nuclear structure function
$G_{b/A}(x_b,\mu^{2})$ in the form

\begin{equation}
G_{b/A}(x_b,\mu^{2}) = A\cdot G_{b/N}(x_b,\mu^{2})\cdot R^A_b(x,\mu^{2}),
\end{equation}
similarly to Ref. \cite{ina} and we take the values of $R^A_b(x,\mu^{2})$
for gluons, valence and sea quarks from Ref. [10]. They are presented in
Fig. 1. The values of $R^A_b(x,\mu^{2})$ in [10] are given for
$x > 10^{-3}$ which is not small enough at high energies. So at $x < 10^{-3}$
we used two  variants of the $R^A_b(x,\mu^{2})$ behaviour for gluon and sea
quark distributions: The first is the constant frozen at $x = 10^{-3}$ 
(dashed-dotted curves for gluons and solid curves for sea quarks in Fig. 1).
The second is the extrapolation as $x^{\beta}$ (dotted curves in Fig. 1)
of the distribution
which gives the main contribution to our cross section, with 
$\beta =$ 0.096 and 0.040 for charm and beauty production respectively (only 
for gluons), and $\beta$ = 0.109, 0.096 and 0.072 for Drell-Yan pair
invariant masses $M^2_{l^+l^-}=$ 5, 25 and 100 GeV$^2$ respectively (only for 
sea quarks). Such behaviour is in qualitative agreement with the results of
\cite{lev}.

In the case of nucleus-nucleus collisions the parton distributions in both
incident nuclei should be written in the form (5). It is necessary to
note that in such a way the main part of the shadowing processes is taken
into account.

\vskip 9 mm

\section{A-DEPENDENCE OF HEAVY FLAVOUR PRODUCTION}

\vskip 2 mm

In the case of charm production we have used the values $m_c$ = 1.5 GeV and
$\mu^2$ = 4 GeV$^2$ ($\mu^2$ = 5 GeV$^2$ for the MRS-A set) and in the case of
beauty production $m_b$ = 5 GeV and $\mu^2$ = $m_b^2$.

The obtained results for $\alpha$ determined from the ratios of heavy quark
production cross section on a gold target and on the proton are presented in
Fig. 2 for the GRV HO and MRS-A sets\footnote{In
all the calculations in this paper
little difference, if any, has been found in the results for $\alpha$ between
these two sets of parton distributions.}
and the two variants of gluon distribution
ratios in the small $x$-region. Here $\sqrt{s_{NN}}$ is the c.m. energy for
the interaction of the incident proton with one target nucleon. One can see
that at fixed target energies the values of $\alpha$ are slightly higher
than unity, which is not in contradiction with the results of Refs. [1,2].
However $\alpha$ decreases with increasing energy and this effect is larger
in the case of charm production than in the case of beauty. One can see
also that the difference between the two variants for $R^A_b(x,\mu^{2})$ at
$x < 10^{-3}$ becomes important only at the highest energies.

We also calculate the values of $\alpha$ for different Feynman-$x$ ($x_F$)
regions using $x_F = x_a - x_b$ at energies $\sqrt{s_{NN}}$ = 39 GeV and
1800 GeV. The results are presented in Fig. 3. At negative and moderate $x_F$
(in the nucleus fragmentation region) the values of $\alpha$ are slightly
higher than unity. However in the case of charm production in the beam
fragmentation region (positive $x_F$) the values of $\alpha$ become
essentially smaller than unity. For beauty production the last effect is
expected only at very high energies.

The obtained results for A-dependences of charm and beauty production in the
symmetric case of gold-gold collisions are presented in Figs. 4 ($\alpha$
as a function of initial energy) and 5 ($\alpha$ as a function of $x_F$ at
two energies). Here $\sqrt{s_{NN}}$ is the c.m. energy for one
nucleon-nucleon interaction. One can see a qualitative agreement with the
case of proton-nucleus collisions (with the difference that the trivial value
of $\alpha$ is here equal to two). Again we can see that the values of
$\alpha$ are dependent on the initial energy and $x_F$ if the energy is high 
enough.

Calculations for charm production in perturbative QCD (including
preequilibrium charm
production from secondary minijet gluons)
can
be found in \cite{wang} for central gold on gold collisions.

\vskip 9 mm

\section{ A-DEPENDENCE OF DRELL-YAN PAIR PRODUCTION}

\vskip 2mm

As said above, the QCD
expression for heavy lepton pair production in a
hadron 1 - hadron 2 collision has the same form as Eq. (1) but with another
matrix element which can be found in Ref. \cite{DY}. Now we have an
additional variable -- the
mass of the produced pair $M_{l^+l^-}$ which plays more
or less the same role as the mass of the heavy quark. However now its value can
be measured experimentally and do not lead to any uncertainty.
We have used the values of QCD scale $\mu^2 = M^2_{l^+l^-}$.

The obtained results for $\alpha$ determined in the same way as in heavy quark
production, i.e., from the ratios of Drell-Yan
pair production cross section in proton-gold and proton-proton
collisions, are presented in Fig. 6 for the GRV HO and MRS-A sets and the two
variants of sea quark distribution ratios in the small
$x$-region\footnote{As is well-known
gluons
are dominant for heavy flavour production, while sea quarks dominate
Drell-Yan production.}. Here again
$\sqrt{s_{NN}}$ is the c.m. energy for the interaction of the incident
proton with one target nucleon. Contrary to the case of heavy flavour
production the values of $\alpha$ are never higher than unity. They
decrease more or less monotonically with increasing energy and this effect
becomes smaller with increasing $l^+l^-$ mass. Again the two sets of parton
distributions give practically identical results for the energy dependence of
$\alpha$ and the difference between the two variants for $R^A_b(x,\mu^{2})$ at
$x < 10^{-3}$ becomes important only at the highest energies. The predicted
$M_{l^+l^-}$ dependence of $\alpha$ for Drell-Yan pair production in
proton-nucleus collisions at different energies is also shown in Fig. 6.

The results of our calculations of $\alpha$ as a function of $x_F$ for
heavy lepton pair production are presented in Fig. 7. The qualitative
picture is similar to the case of heavy flavour production; the
numerical difference of our predictions from the value $\alpha = 1$ is here
more significant for the case of a not very large mass of the lepton pair.

Predictions of A-dependence for Drell-Yan pair production cross section
in symmetric nucleus-nucleus collisions are shown in Figs. 8 and 9. As for
the case of proton-nucleus collisions, all the effects are
qualitatively the same but numerically larger than in the case of heavy
flavour production. Our results are in agreement with the
calculations of \cite{GGRV}.

\vskip 9 mm

\section{CONCLUSIONS}

\vskip 2 mm

In this paper
we calculate the A-dependence of charm and beauty as well as Drell-Yan pair
production using standard QCD formulas and accounting for the difference of
parton distributions in free and bound nucleons. If one parametrize these
cross sections as $\sigma \sim A^{\alpha}$, the value of $\alpha$ is slightly
different from unity at the available energies. For the case of heavy
flavour production at comparatively low energies the obtained values of
$\alpha$ are a little higher than unity. This should be connected with some
nucleon-nucleon correlations which change the large-$x$  parton
distributions. In the case of Drell-Yan pair production such effect is
absent because of the different relations in the contributions of valence
quarks, sea
quarks and gluons.

At higher energies the values of $\alpha$ decrease and become smaller than
unity. At $\sqrt{s_{NN}}$ = 1800 GeV we expect a value of $\alpha \sim 0.95$
for charm and low-mass Drell-Yan pair production.
The decrease of the ratio $R^A_b(x,\mu^{2})$, which results in a decrease of
$\alpha$, can be connected with the effects of parton density saturation
\cite{glr} which in heavy nuclei occur at $x$-values higher than in the
proton.

If we consider two small and different values of $x_a$ and $x_b$ in Eq. (1)
for the case of proton-nucleus interaction, it is clear that the contribution 
to the inclusive cross section from the region $x_a < x_b$ should be larger 
than the mirror contribution ($x_a > x_b$) because the value of the ratio 
$R^A_b(x,\mu^{2})$ in the first case is larger. It means that heavy quark and 
Drell-Yan pairs will be produced preferably in the nucleus fragmentation 
hemisphere, i.e., asymmetrically, which is quite usual and has been confirmed 
experimentally in the case of light quark production. From Figs. 3 and 7 it is 
clear that charm and low-mass Drell-Yan pair production on nuclear targets at 
LHC energies will give important information on the nuclear shadowing of the 
structure functions at small $x$.

Besides,
the experimental measurements of the effects predicted for different hard
interactions should allow (in principle) to control the validity of the
conventional extraction of parton distributions from the experimental DIS
data.

Let us repeat again that almost all nuclear shadowing corrections are accounted
for in our calculations because they contribute to the $R^A_b(x,\mu^{2})$
ratios, which have been extracted from the experimental data. On the other hand
these shadowing corrections are not numerically large in the considered hard
processes. So we can assume that the corrections which are not accounted for
do not change significantly the obtained results.


In conclusion we express our gratitude to M.A.Braun for useful discussions and 
both to K.J.Eskola and to I.Sarcevic for sending us their numerical results. 
We thank the Direcci\'on General de Pol\'{\i}tica Cient\'{\i}fica and the 
CICYT of Spain for financial support under contract AEN96-1673. C.A.S. also 
thanks the Xunta de Galicia for financial support. The paper was supported in 
part by INTAS grant 93-0079.

\newpage

\noindent{\Large \bf Figure captions}
\vspace{0.5cm}

\noindent{\bf Fig. 1.}
Functions $R^A_G(x,\mu^{2})$, $R^A_V(x,\mu^{2})$ and
$R^A_S(x,\mu^{2})$, which determine the ratios of the distributions for
protons in the nucleus versus free protons, for gluons (dashed-dotted and
dotted
curves, see text), valence quarks (dashed curves) and sea quarks (solid and
dotted
curves, see text)
respectively, for $\mu^2 =$ 5 GeV$^2$ (upper figure) and $\mu^2 = m_b^2$ (lower
figure).

\noindent{\bf Fig. 2.}
Energy dependence of $\alpha$ for charm and beauty
production in proton-gold collisions for GRV HO (solid and dashed curves) and
MRS-A (dotted and dashed-dotted curves) structure
functions and using extrapolated (dashed and dashed-dotted curves) and frozen at
$x= 10^{-3}$ (solid and dotted curves) ratios of gluon distributions.

\noindent{\bf Fig. 3.}
Feynman-$x$ dependence of $\alpha$ for charm
and beauty production in proton-gold collisions at $\sqrt{s_{NN}}$ = 39 GeV
(upper figure) and 1800 GeV (lower figure)
for GRV HO and MSR-A structure functions and using
extrapolated and frozen at $x = 10^{-3}$
ratios of gluon distributions
(with the same conventions as in Fig. 2).

\noindent{\bf Fig. 4.}
Energy dependence of $\alpha$ for charm and beauty
production in gold-gold collisions for GRV HO and MRS-A structure functions
and using extrapolated and frozen at $x = 10^{-3}$
ratios of gluon distributions
(with the same conventions as in Fig. 2).

\noindent{\bf Fig. 5.}
Feynman-$x$ dependence of $\alpha$ values for charm
and beauty production in gold-gold collisions at $\sqrt{s_{NN}}$ = 39 GeV
(upper figure) and 1800 GeV (lower figure)
for GRV HO and MSR-A structure functions and using
extrapolated and frozen at  $x = 10^{-3}$
ratios of gluon distributions
(with the same conventions as in Fig. 2).

\noindent{\bf Fig. 6.}
Mass (upper figure)
and energy (lower figure)
dependence of $\alpha$ for
Drell-Yan pair production in proton-gold collisions for GRV HO and MRS-A
structure functions and using extrapolated and
frozen at $x = 10^{-3}$ ratios of sea quark distributions
(with the same conventions as in Fig. 2).

\noindent{\bf Fig. 7.}
Feynman-$x$ dependence of $\alpha$ values for heavy
lepton pair production in proton-gold collisions at $\sqrt{s_{NN}}$ = 39 GeV
(upper curves in each figure) and 1800 GeV (lower curves in each figure)
and different masses
for GRV HO and MSR-A structure functions and using
extrapolated and frozen at  $x = 10^{-3}$
ratios of sea quark distributions
(with the same conventions as in Fig. 2). Note that at $\sqrt{s_{NN}}$ = 39 GeV
all curves coincide.

\noindent{\bf Fig. 8.}
Mass (upper figure)
and energy (lower figure)
dependence of $\alpha$ for Drell-Yan pair
production in gold-gold collisions for GRV HO and MRS-A structure functions
and using extrapolated and frozen at $x = 10^{-3}$
ratios of sea quark distributions
(with the same conventions as in Fig. 2).

\noindent{\bf Fig. 9.}
Feynman-$x$ dependence of $\alpha$ values for heavy
lepton pair production in gold-gold collisions at $\sqrt{s_{NN}}$ = 39 GeV
(upper curves in each figure) and 1800 GeV (lower curves in each figure)
and different masses
for GRV HO and MSR-A structure functions and using
extrapolated and frozen at  $x = 10^{-3}$
ratios of sea quark distributions
(with the same conventions as in Fig. 2). Note that at $\sqrt{s_{NN}}$ = 39 GeV
all curves coincide.

\end {document}
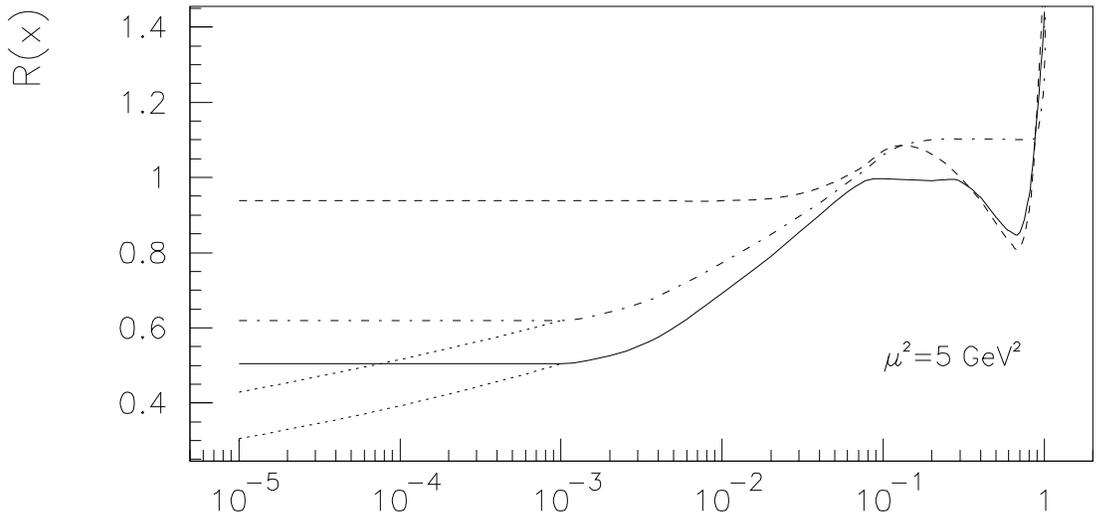
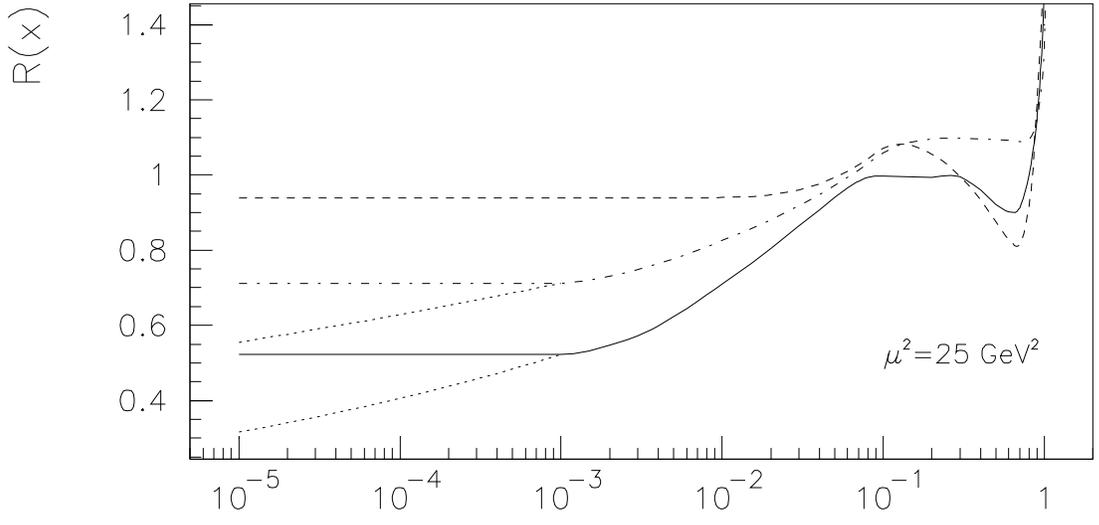
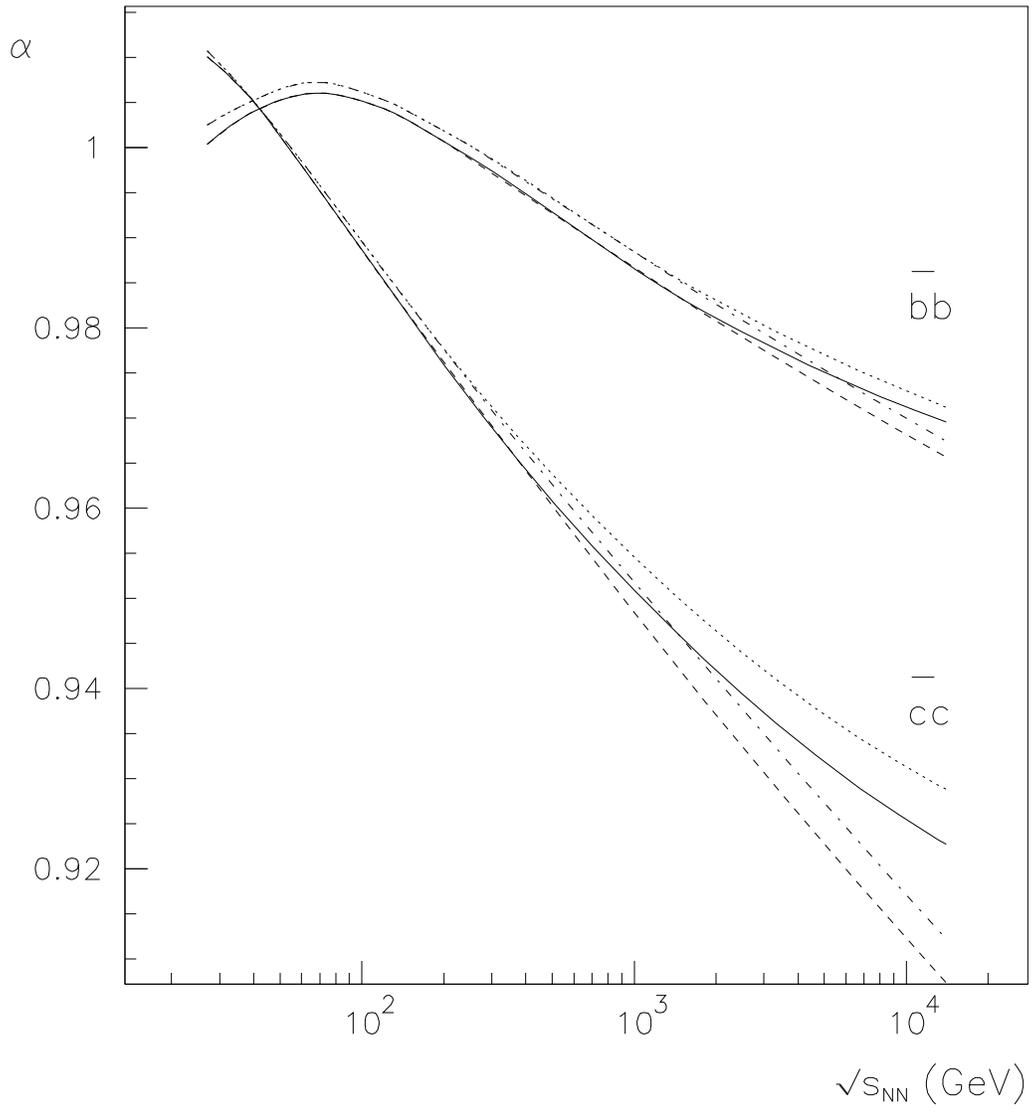
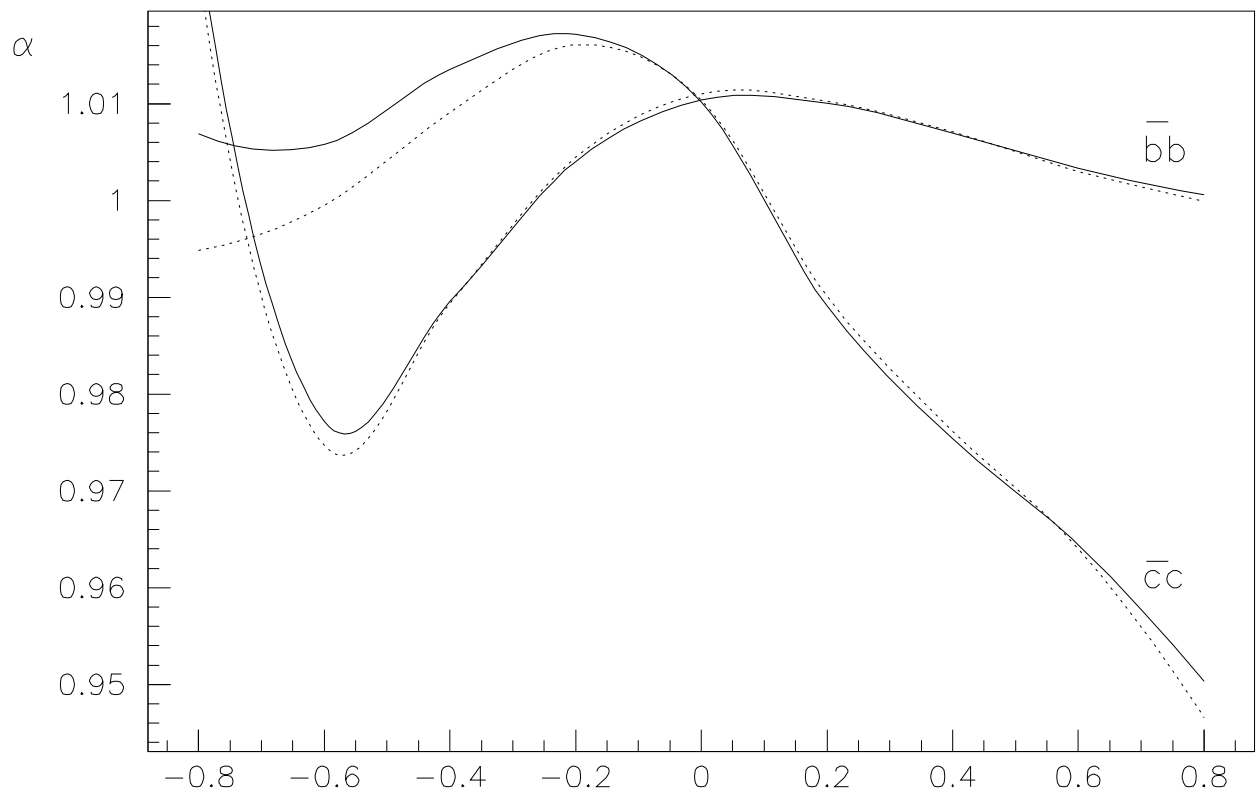
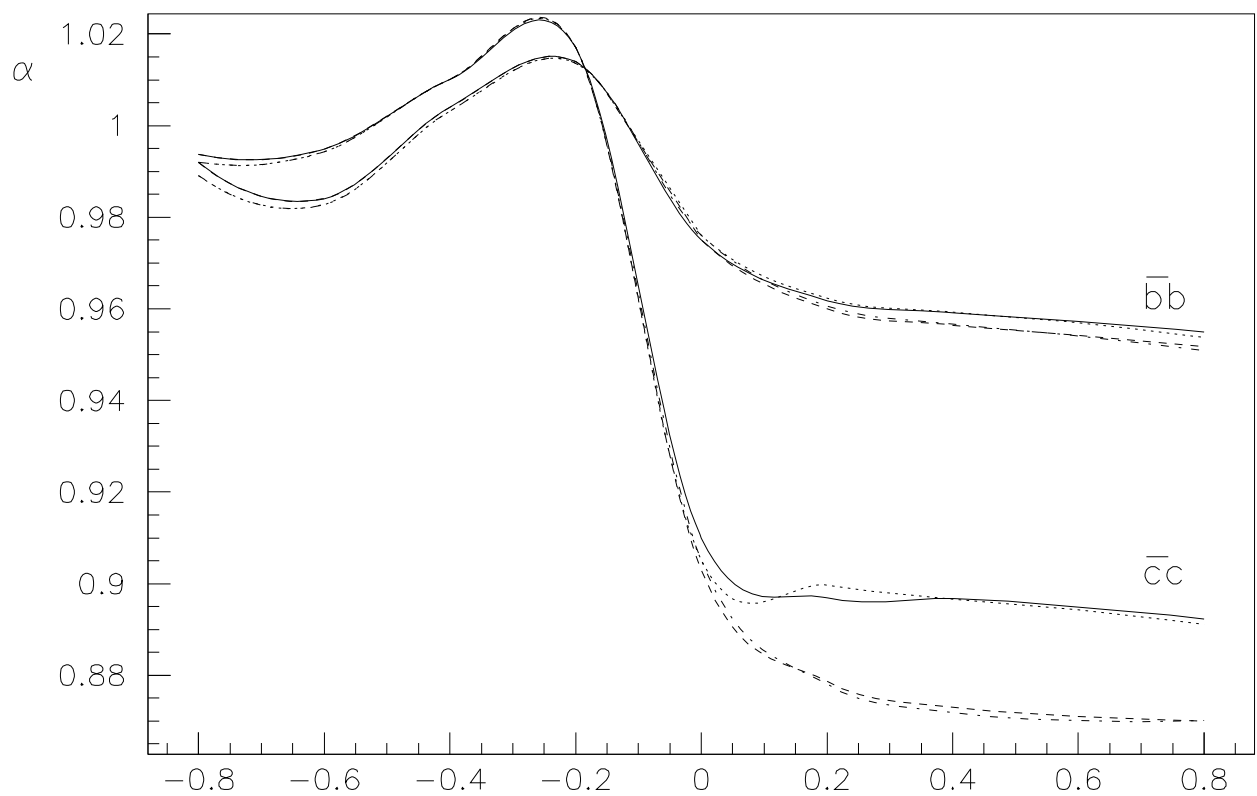
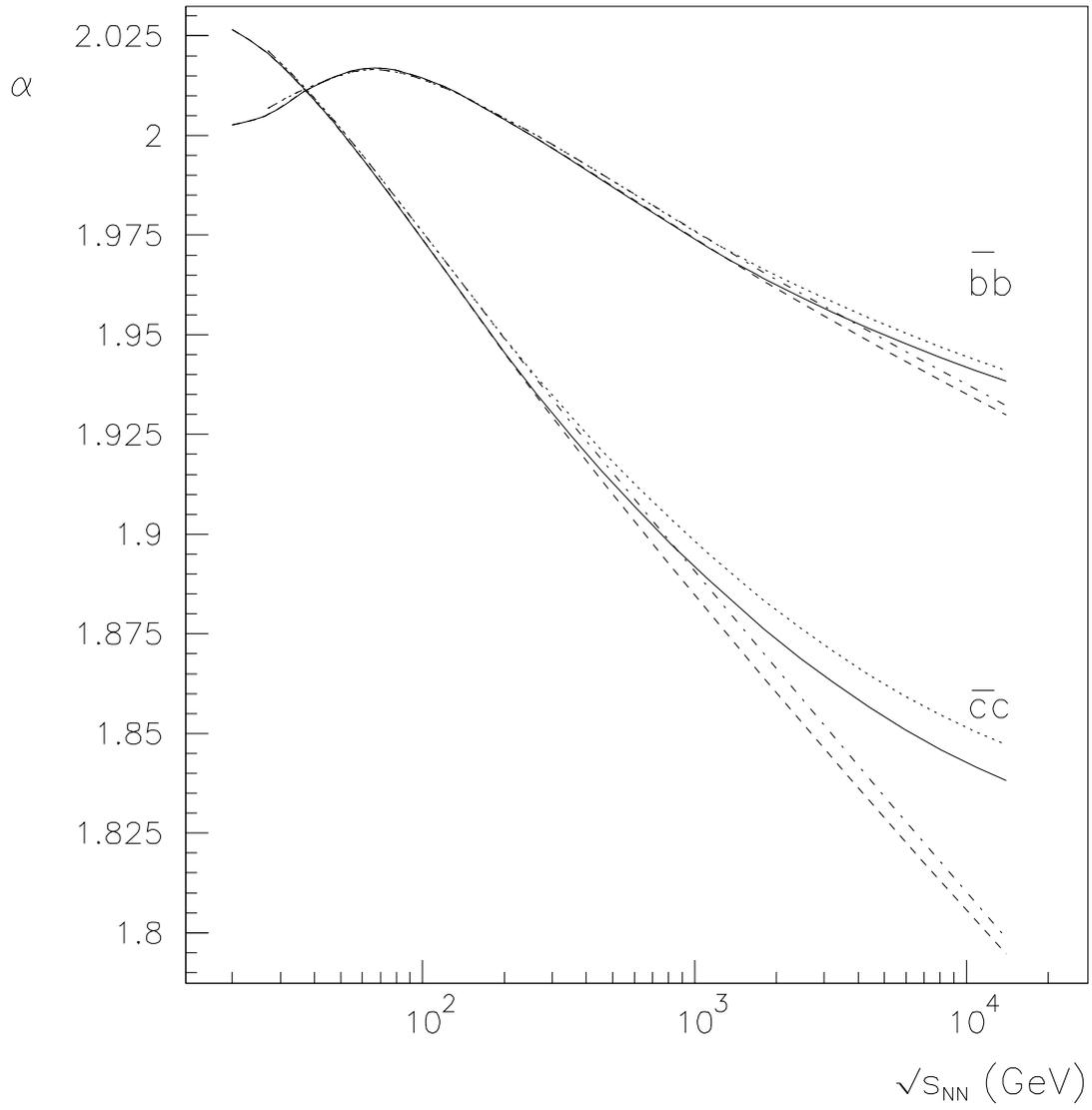
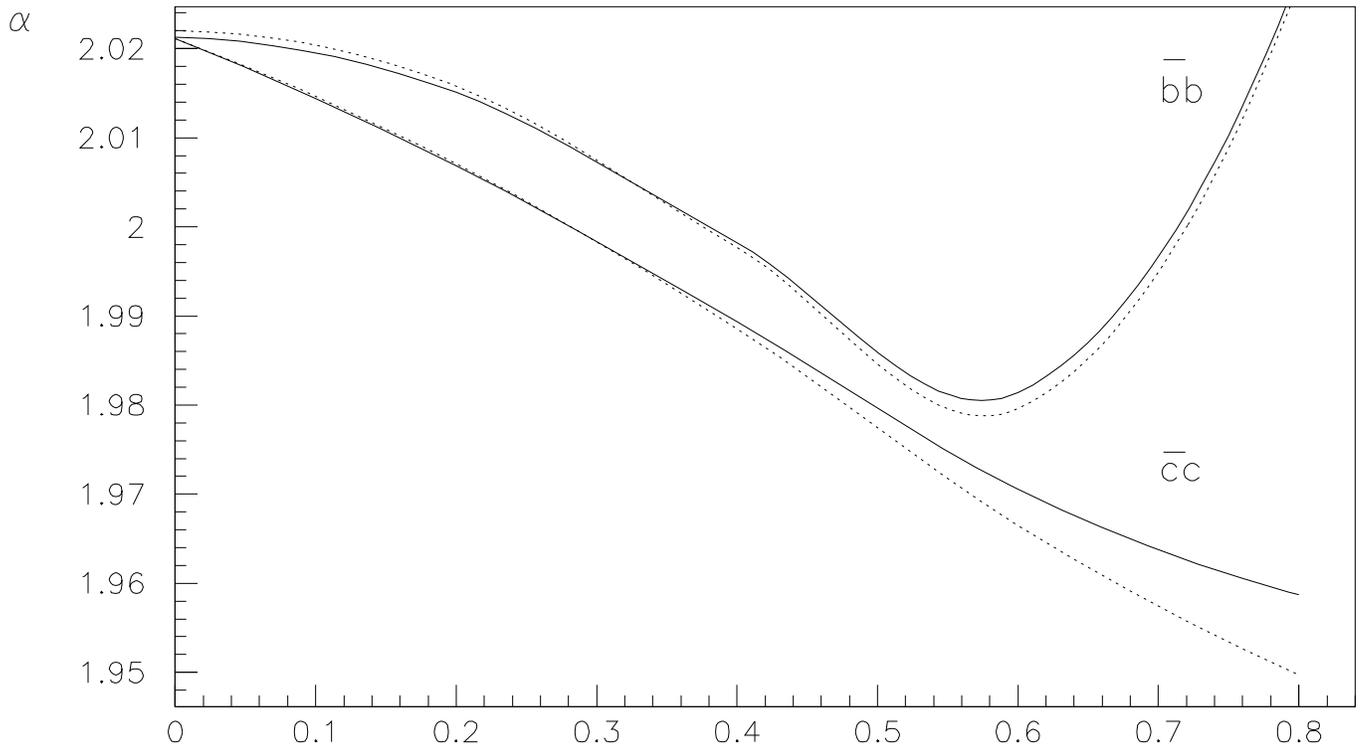
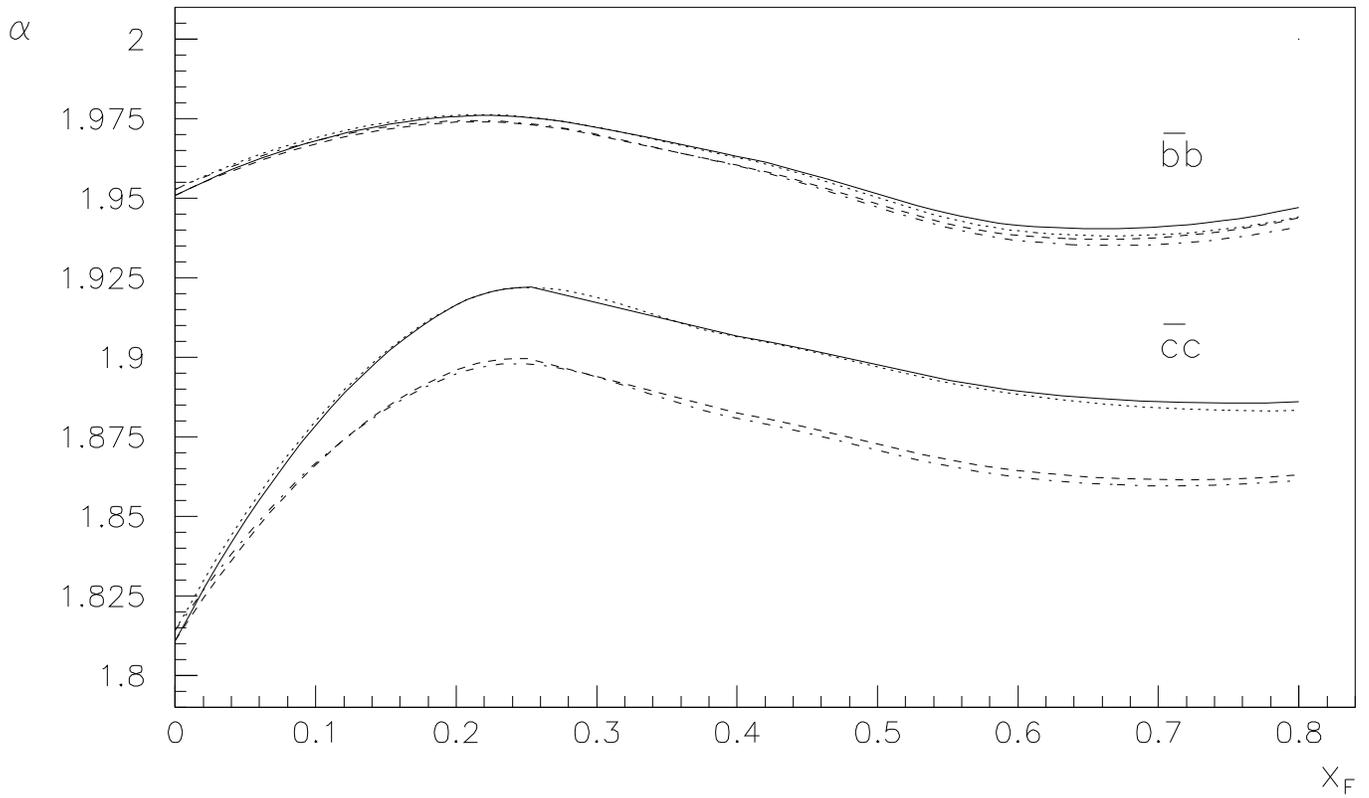
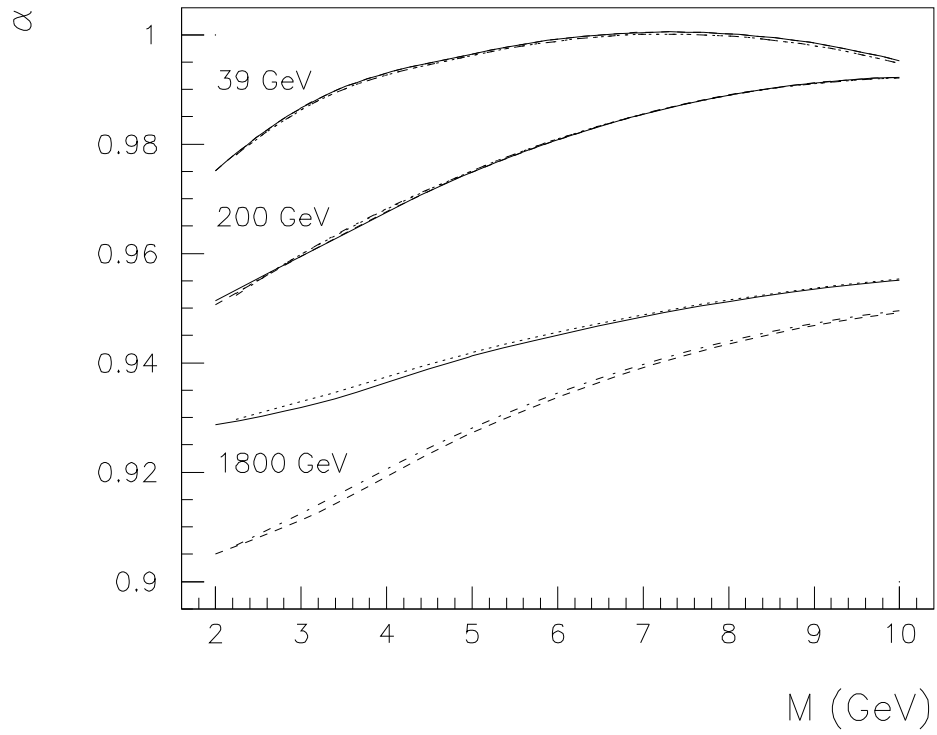
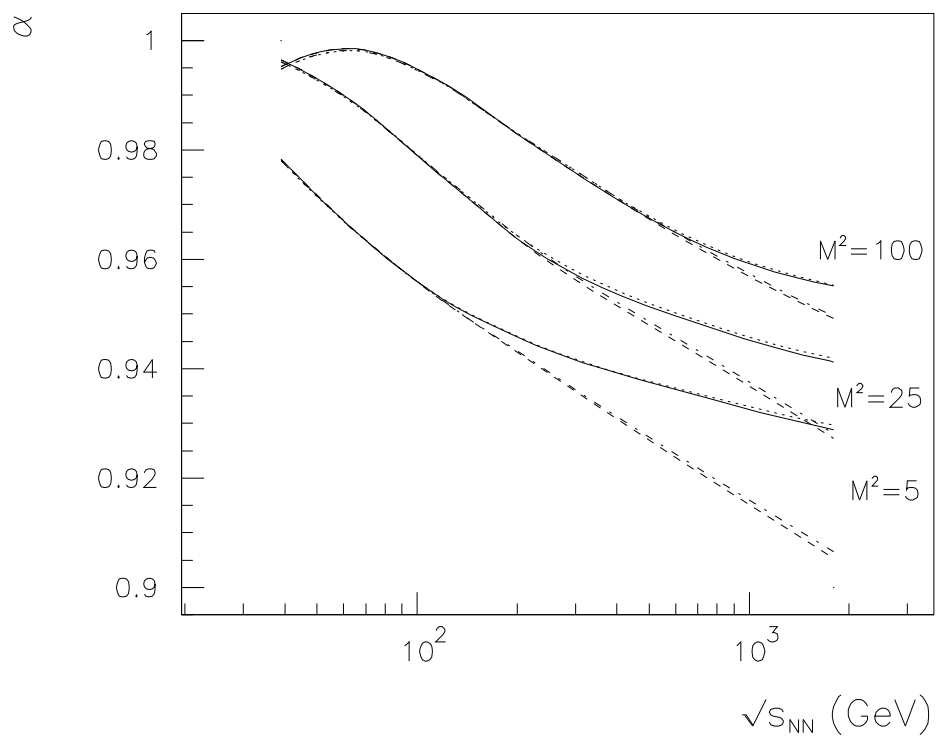
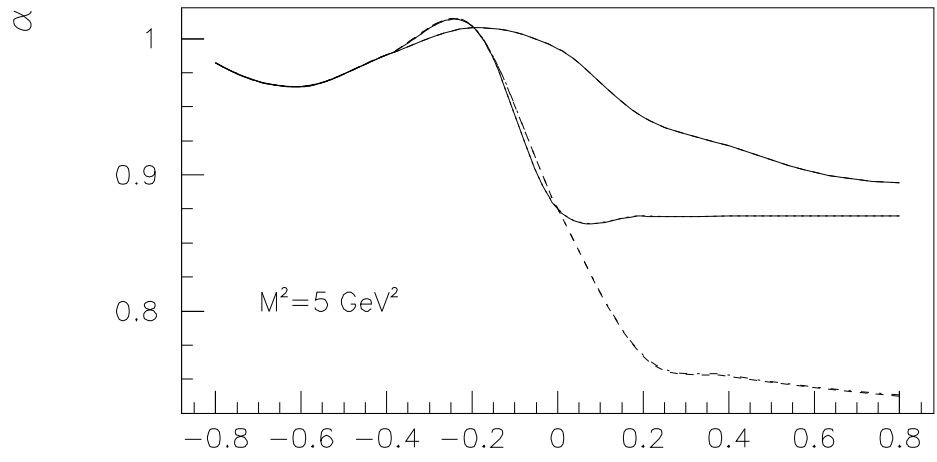
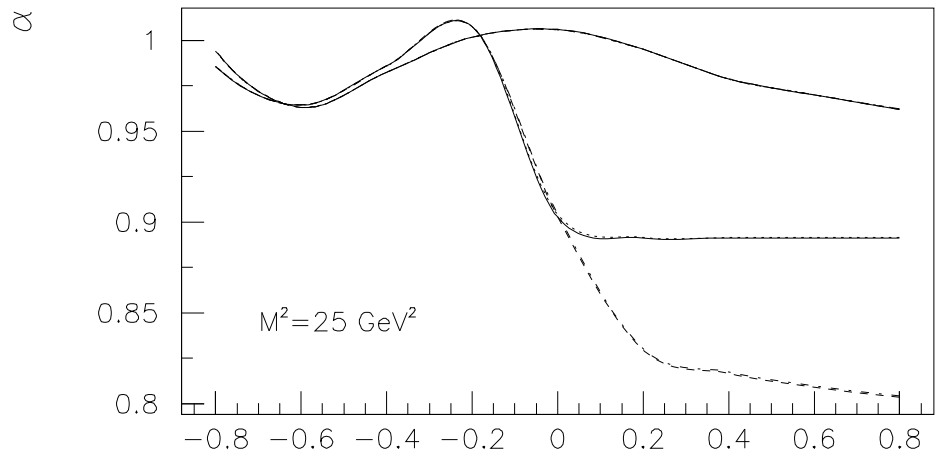
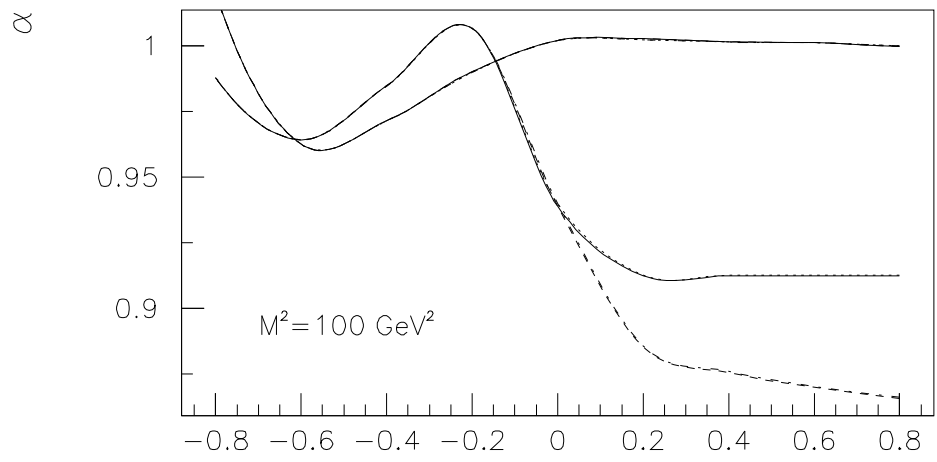
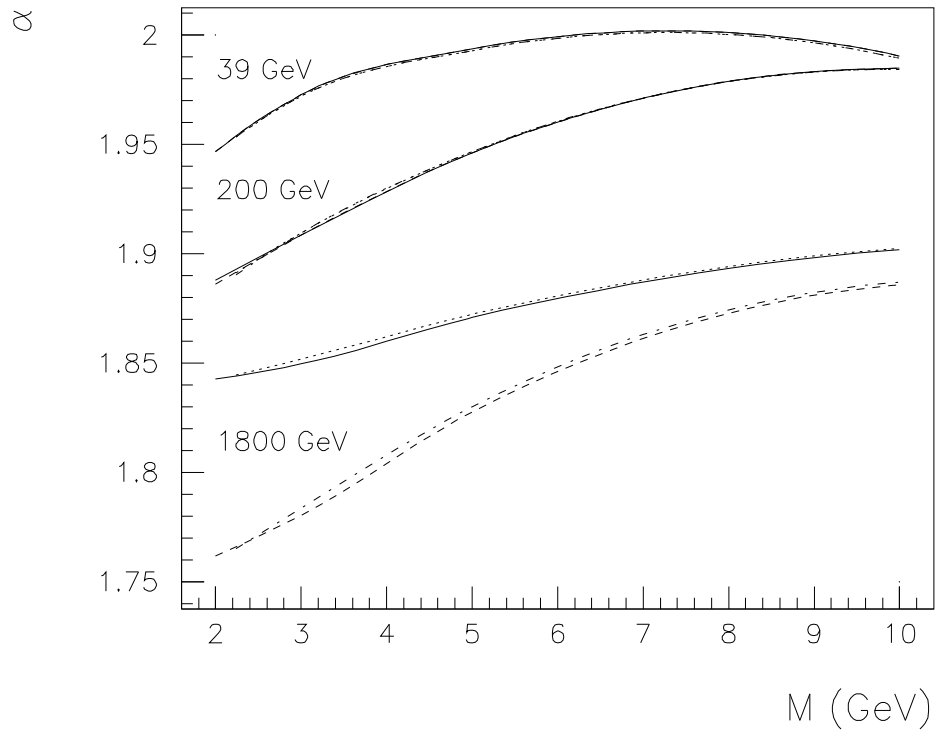
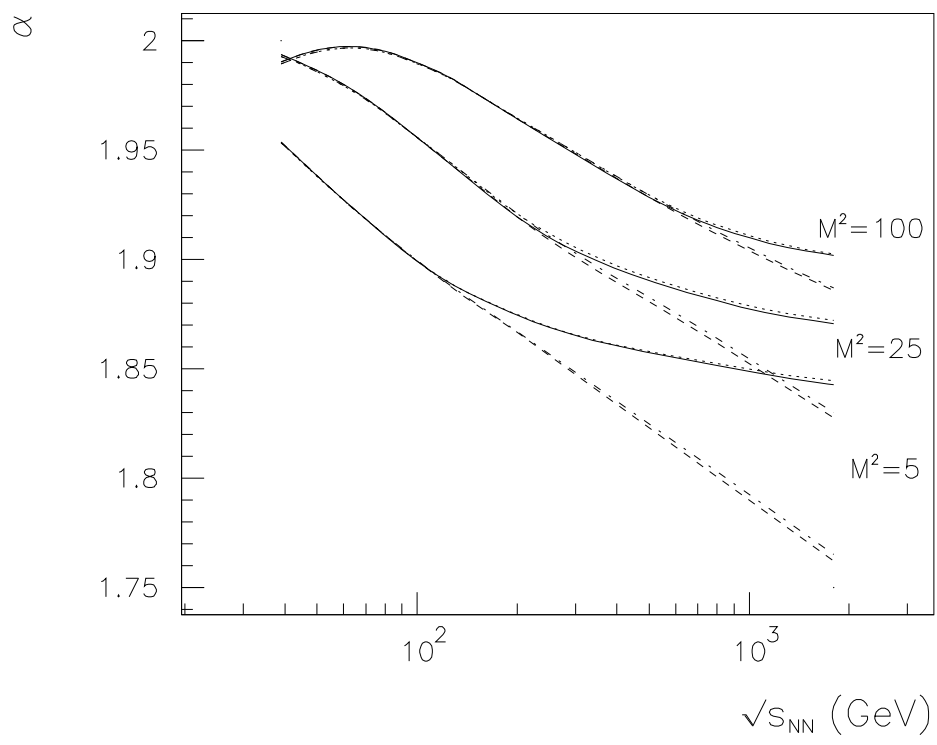
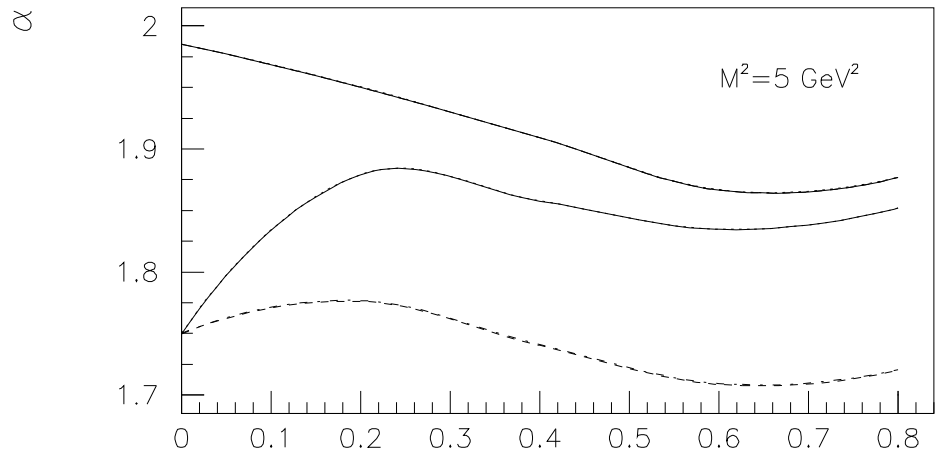
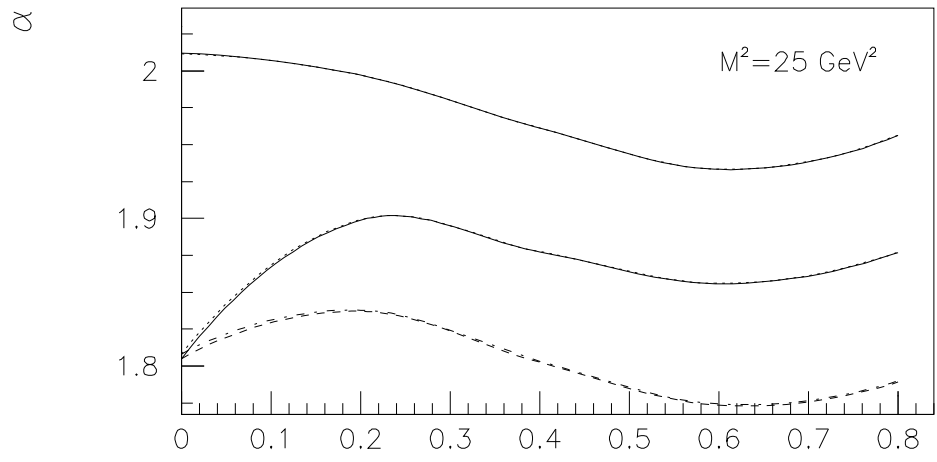
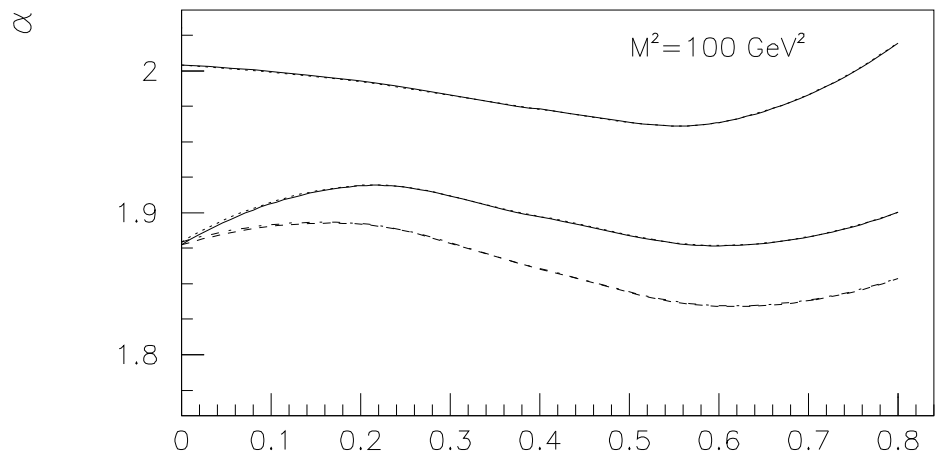
